\documentclass[letterpaper,titlepage,11pt]{article}
\usepackage[colorlinks=true,urlcolor=blue,citecolor=magenta]{hyperref}
\usepackage{amssymb,amsmath,amsfonts}
\usepackage{epsfig}
\usepackage{graphicx}

\def\clock{{\count0=\time
           \divide\count0 60
           \ifnum\count0<10 0\fi\the\count0
           \multiply\count0 -60 \advance\count0 \time
           :\ifnum\count0<10 0\fi \the\count0
         }}
\newcommand{\timestamp}{{\small\vbox{\hbox{\tt\jobname.tex}
\hbox{\the\day/\the\month/\the\year, \clock}}}}

\newcommand{\be}{\begin{eqnarray}}
\newcommand{\ee}{\end{eqnarray}}
\newcommand{\beq}{\begin{equation}}
\newcommand{\eeq}{\end{equation}}

\newcommand{\beqa}{\begin{eqnarray}}
\newcommand{\eeqa}{\end{eqnarray}}

\let\oldsqrt\sqrt
\def\sqrt{\mathpalette\DHLhksqrt}
\def\DHLhksqrt#1#2{%
\setbox0=\hbox{$#1\oldsqrt{#2\,}$}\dimen0=\ht0
\advance\dimen0-0.2\ht0
\setbox2=\hbox{\vrule height\ht0 depth -\dimen0}%
{\box0\lower0.4pt\box2}}

\numberwithin{equation}{section}
\begin{document}

\begin{titlepage}

\topmargin=-1true mm

 \vskip 3 cm

\centerline{\Huge \bf Black Branes as Piezoelectrics}

\vskip 1.5cm

\centerline{\large {\bf Jay Armas},  {\bf Jakob Gath} and {\bf Niels A. Obers}.}

\vskip 1cm

\begin{center}
\sl The Niels Bohr Institute, Copenhagen University  \\
\sl  Blegdamsvej 17, DK-2100 Copenhagen \O , Denmark
\end{center}
\vskip 0.2cm

\centerline{\small\tt jay@nbi.dk, gath@nbi.dk, obers@nbi.dk}

\vskip 3cm \centerline{\bf Abstract} \vskip 0.2cm \noindent
We find a realization of linear electroelasticity theory in gravitational physics by uncovering a new response coefficient of charged black branes, exhibiting their piezoelectric behavior. Taking charged dilatonic black strings as an example and using the blackfold approach we measure their elastic and piezolectric moduli. We also use our results to draw predictions about the equilibrium condition of charged dilatonic black rings in dimensions higher than six.

\end{titlepage}

\tableofcontents

\section{Introduction}
The dynamics of black branes is governed by dissipative fluid mechanics in the regime of long-wavelength perturbations. This has been shown in the context of the AdS/CFT correspondence where the fluid lives on the boundary \cite{Bhattacharyya:2008jc} and for asymptotically flat space-times using holographically inspired techniques \cite{Emparan:2009cs, Emparan:2009at, Camps:2010br}. The dissipative character of these fluid flows allows for the measurement of transport coefficients such as bulk and shear viscosities by evaluating the boundary stress-energy tensor in a derivative expansion. These transport coefficients have been used to draw predictions about the viscosity to entropy ratio of strongly coupled plasmas in holographic setups and by extrapolation in real world physical models \cite{Policastro:2001yc}. In the context of asymptotically flat space, knowledge of the viscosity coefficients is enough to predict the onset of the Gregory-Laflamme instability with higher accuracy for larger brane co-dimension, indicating that the full non-linearity of Einstein's equations is effectively reduced to the simple physical system of viscous fluid flows \cite{Camps:2010br}.

The character of the perturbations that has received the greatest deal of attention in the literature is of the intrinsic type, in which small deformations of the black brane geometry are applied along the space-time boundary or the brane worldvolume directions. However, perturbations which are of extrinsic nature have been considered within the framework of the blackfold approach \cite{Emparan:2007wm, Emparan:2009cs, Emparan:2009at}, where strains are induced by bending black brane geometries along transverse directions to the brane. In its simplest setting, the blackfold approach is an effective theory governing the dynamics of black $p$-branes with thickness $r_0$ wrapped over a submanifold $\mathcal{W}_{p+1}$ with characteristic length scale $R\gg r_0$ in the ambient space-time. The set of equations that describe the blackfold effective theory, directly derivable from Einstein equations \cite{Camps:2012hw}, is a generalization of usual fluid mechanics in which the fluid is confined to a dynamical surface which acts as a boundary for the near-horizon geometry of a black $p$-brane. Extrinsic perturbations can be accounted for through a derivative expansion in the parameter $\varepsilon\equiv \left(r_0/R\right)$ which takes into consideration the finite thickness effects of the brane geometry. In stark contrast with viscous intrinsic corrections where the dynamics is time-dependent, these deformations have been applied to stationary fluid configurations leading to the uncovering of new response coefficients \cite{Armas:2011uf} of uncharged black branes that can be interpreted as elastic moduli. This has provided the first measurement of a general relativistic generalization of the Young modulus. 

The application of the blackfold approach, which can be considered as a generalization of the fluid/gravity correspondence, has led to a broader connection between gravitational physics and material science exhibiting the fluid description along worldvolume or boundary directions on one side and the elastic solid description along transverse directions of thin black brane geometries on the other side. This linkage is still in its early stages of development and the purpose of this letter is to search for more general connections between gravity and long-wavelength physics, and in particular probe the rich landscape of possible response coefficients of black branes. In this context we present and unravel the electroelastic behavior of charged black strings in Einstein-Maxwell-Dilaton (EMD) theories by explicitly measuring their piezoelectric moduli - a relativistic notion of the response coefficients encountered in real world piezoelectrics - and hence draw a novel parallel between gravitational physics and electroelasticity theory.

\section{Equations of Motion}
Hydrodynamical perturbations of black branes are sourced by a monopolar distribution of stress-energy tensor which to leading order is of the perfect fluid form and receives  corrections order-by-order in a boundary derivative expansion. On the other hand, extrinsic elastic deformations are introduced by considering finite thickness corrections to the brane geometry which are captured by a multipolar expansion of the stress-energy tensor. This is done by correcting the stress-energy tensor in a Dirac-delta function series:
\beq \label{stress_tensor}
\hat{T}^{\mu\nu}(x^{\lambda})=\int_{\mathcal{W}_{p+1}}\!\!\!\!\! dV\Bigg[\frac{T^{\mu\nu}_{(0)}(\sigma^a)}{\sqrt{-g}}\delta^{(D)}(x^{\lambda}-X^{\lambda}(\sigma^{a}))-\nabla_{\rho}\left(\frac{T^{\mu\nu\rho}_{(1)}(\sigma^{a})}{\sqrt{-g}}\delta^{(D)}(x^{\lambda}-X^{\lambda}(\sigma^{a}))\right)+...\Bigg]~,
\eeq
where $dV=d^{p+1}\sigma\sqrt{-\gamma}$. Our conventions are as follows: $\mu,\nu$ are space-time indices, $a,b$ are $p$-brane worldvolume indices and $X^{\mu}(\sigma^{a})$ denotes the brane embedding functions. Furthermore, $g_{\mu\nu}$ is the background metric and $\gamma_{ab}=g_{\mu\nu}u^{\mu}_a u^{\nu}_b$ with $u^{\mu}_{a}\equiv \partial_a X^{\mu}$ is the induced metric on the brane worldvolume. The Dirac-delta function has the role of localizing the source in the space-time region $x^{\mu}=X^{\mu}(\sigma^a)$ when dipolar terms are absent, otherwise these introduce an ambiguity in the position of the worldvolume surface within a finite size region of thickness $r_0$. This is parametrized by `extra symmetry 2' \cite{Vasilic:2007wp} which transforms $T^{\mu\nu}_{(0)}$ and $T^{\mu\nu\rho}_{(1)}$ under a $\mathcal{O}(\varepsilon)$ displacement of the worldvolume location $X^{\rho}(\sigma^{a})\to X^{\rho}(\sigma^{a})+\epsilon^{\rho}(\sigma^a)$.

In Eq.~\eqref{stress_tensor} the tensor $T^{\mu\nu}_{(0)}(\sigma^{a})$ represents the monopolar structure of the $(p+1)$-dimensional source and receives dissipative corrections when hydrodynamic fluctuations are considered, while the extra structure $T^{\mu\nu\rho}_{(1)}(\sigma^{a})$ encodes the dipolar character of the distribution of stress-energy. Assuming the brane not to backreact onto the background space-time, the equations of motion in the absence of external forces follow from stress-energy conservation
\beq \label{stress_conservation}
\nabla_{\nu}\hat{T}^{\mu\nu}=0~.
\eeq
This gives rise to a worldvolume effective theory that describes the dynamics of fluid configurations living on dynamical surfaces \cite{Emparan:2009at, Armas:2011uf}. Eq.~\eqref{stress_conservation} is valid also for charged (dilatonic) branes as long as no couplings to external background fields are present \cite{Emparan:2011hg, Caldarelli:2010xz, Grignani:2010xm}.

Finite size dipole corrections to the stress-energy tensor are well known to capture intrinsic spin when considering the motion of spinning point particles, while as shown in \cite{Armas:2011uf}, they can also effectively describe worldvolume stress-energy dipoles induced by the bending of black branes. More precisely, using the orthogonal projector ${\perp^{\mu}}_{\nu}={\delta^{\mu}}_{\nu}-u^{\mu}_{a}u^{a}_{\nu}$, the 3-index tensor structure $T^{\mu\nu\rho}_{(1)}$ can be decomposed as follows
\beq
T^{\mu\nu\rho}_{(1)}=u^{(\mu}_{b}j^{\nu)\rho b}+u^{\mu}_{a}u^{\nu}_{b}d^{ab\rho}+u^{\rho}_{a}T^{\mu\nu a}_{(1)}~.
\eeq
Here $j^{\nu\rho b}\equiv {\perp^{\nu}}_{\mu}{\perp^{\rho}}_{\lambda}u^{b}_{\sigma}T^{\mu\lambda\sigma}_{(1)}$, satisfying $j^{\nu\rho b}=j^{[\nu\rho]b}$, encodes the intrinsic angular momenta, $d^{ab\rho}\equiv u^{a}_{\mu}u^{b}_{\nu}{\perp^{\rho}}_{\lambda}T^{\mu\nu\lambda}_{(1)}$, with property $d^{ab\rho}=d^{(ab)\rho}$, describes the dipole sources of worldvolume stress-energy and the components $T^{\mu\nu a}_{(1)}$ are pure gauge and can be set to zero using the general covariance of \eqref{stress_tensor}.

Electrically charged fluid configurations have also been considered in the fluid/gravity correspondence when dealing with hydrodynamic fluctuations, allowing for the computation of conductivities \cite{Erdmenger:2008rm,Banerjee:2008th}. Besides being described by a stress-energy tensor, these are characterized by an additional monopole source of electric current. Here we are interested in black branes that are electrically charged under a 2-form field strength $F^{\mu\nu}(x^{\alpha})$ which develop worldvolume electric dipoles due to the action of bending. In direct analogy with classical electrodynamics we expand the electric current in a Dirac-delta function series:
\beq \label{current}
\hat{J}^{\mu}(x^{\lambda})=\int_{\mathcal{W}_{p+1}}\!\!\!\!\! dV\Bigg[\frac{J^{\mu}_{(0)}(\sigma^a)}{\sqrt{-g}}\delta^{(D)}(x^{\lambda}-X^{\lambda}(\sigma^{a})) -\nabla_{\rho}\left(\frac{J^{\mu\rho}_{(1)}(\sigma^{a})}{\sqrt{-g}}\delta^{(D)}(x^{\lambda}-X^{\lambda}(\sigma^{a}))\right)+...\Bigg]~.
\eeq
In parallel with \eqref{stress_tensor}, this displays general covariance via `extra symmetry 1', $\delta_1 J^{\mu\rho}=-\epsilon^{\mu a}u^{\rho}_{a}$ and $\delta_1 J^{\mu}=-\nabla_{a}\epsilon^{\mu a}$, and has the same ambiguity as the stress-energy tensor under small displacements of the embedding surface expressed by `extra symmetry 2': 
\beq
\begin{split}
\delta_2J^{\mu}_{(0)}&=-J_{(0)}^{\mu}u^{a}_\rho \nabla_{a}\epsilon^{\rho}-J^{\lambda}_{(0)}\Gamma^{\mu}_{\rho\lambda}\epsilon^{\rho}\, ,\\ 
\delta_2 J^{\mu\rho}_{(1)}&=-J^{\mu}_{(0)}\epsilon^{\rho}~.
\end{split}
\eeq
In the probe approximation, Eq.~\eqref{stress_conservation} is now supplemented  with current conservation:
\beq \label{current_conservation}
\nabla_{\mu}\hat{J}^{\mu}=0~.
\eeq
Following the same method as in \cite{Vasilic:2007wp}, we can solve Eq.~\eqref{current_conservation} decomposing the monopole part of the current as
\beq
J_{(0)}^{\mu}=u^{\mu}_{a}J^{a}_{(0)}+J_{\perp}^{\mu}, \quad J_{\perp}^{\mu}={\perp^{\mu}}_{\lambda}J^{\lambda}_{(0)}~,
\eeq
and the 2-index structure $J^{\mu\rho}_{(1)}$ as
\beq \label{decomp_current}
J^{\mu\rho}_{(1)}=m^{[\mu\rho]}+u^{\mu}_{a}p^{a\rho}+J^{\mu a}_{(1)}u_{a}^{\rho}~,
\eeq
where we have used `extra symmetry 1' to gauge away some of the components. Here $m^{[\mu\rho]}={\perp^{[\mu}}_{\nu}{\perp^{\rho]}}_{\lambda}J^{\nu\lambda}_{(1)}$ is an additional contribution to the electric current due to the motion in transverse directions, $p^{a\rho}=u^{a}_{\mu}{\perp^{\rho}}_{\nu}J^{\mu\nu}_{(1)}$ is the electric dipole moment while the components $J^{\mu a}_{(1)}$ are pure gauge and can be set to zero. The extra components of the monopole part of the current are not independent and can be related to the dipolar part by the relation $J_{\perp}^{\mu}={\perp^{\mu}}_{\lambda}\nabla_{a}\left(2J^{(\lambda a)}_{(1)}-J^{(ab)}_{(1)}u_{b}^{\lambda}\right)$. Consequently, current conservation \eqref{current_conservation} results in the worldvolume conservation equation
\beq \label{curren_conservation}
\nabla_{a}\left(\mathcal{J}^{a}+p^{b\rho}\nabla_{b}u^{a}_{\rho}\right)=0~,
\eeq
where we have defined the worldvolume tensor $\mathcal{J}^{a}=J^{a}_{(0)}-\nabla_{b}J^{(ab)}_{(1)}$. In the case of worldvolumes with boundaries Eq.~\eqref{current_conservation} must be supplement by additional boundary conditions \cite{Armas:2012}. When applied to the special case $p=0$, these equations reduce to those derived for the charged spinning point particle \cite{Dixon:1974, Souriou:1974}, though the complete decomposition \eqref{decomp_current}, crucial for the physical interpretation that follows, has not been considered in the literature.

\subsection*{Electroelasticity of Black Branes}
The dipole moment of worldvolume stress-energy $d^{ab\rho}$ is not a priori constrained. Under the expectation that bent black branes will behave like elastic solids we assume the following relation between the dipole moment and the strain in transverse directions ${K_{cd}}^{\rho}\equiv\nabla_{c}u_{d}^{\rho}$ as argued in \cite{Armas:2011uf}:
\beq \label{dipole_stress}
d^{ab\rho}=\tilde{Y}^{abcd}{K_{cd}}^{\rho}~,
\eeq
where $\tilde{Y}^{abcd}$ are the elastic moduli that characterize the brane response to the bending. The assumption \eqref{dipole_stress} has a direct analogy with Hookean classical elasticity theory. In the following we make a further assumption, namely that a bent charged brane will behave like a piezoelectric material in the manner dictated by linear electroelasticity theory:
\beq \label{dipole_current}
p^{a\rho}=\tilde{\kappa}^{abc}{K_{bc}}^{\rho}~,
\eeq
where $\tilde{\kappa}^{abc}$ are the piezoelectric moduli that capture the response of the bent charged material due to electroelastic deformations. In the case of a point particle, the components $p^{a\rho}$ can be gauged away using the `extra symmetry 2'.  To further motivate this interpretation we consider the equations of motion for a spinless charged string bent over a circle of radius $R$ in Minkowski space-time:
\beq \label{eom_blackstring}
D_{a}T^{ab}=0, \quad T^{ab}{K_{ab}}^{\rho}=0, \quad D_{a}J^{a}=0~,
\eeq
where $T^{ab}=T^{ab}_{(0)}+\tau^{ab}_{(1)}$ and $J^{a}=J^{a}_{(0)}+\Upsilon^{a}_{(1)}$ are the effective worldvolume stress-energy tensor and current respectively. The linear electroelastic corrections are
\beq
\tau_{(1)}^{ab}=\tilde{Y}^{abcd}{K_{cd}}^{\rho}K_{\rho}~, \quad \Upsilon^{a}_{(1)}=\tilde{\kappa}^{abc}{K_{bc}}^{\rho}K_{\rho}~,
\eeq
which arise along the direction set by the mean extrinsic curvature $K^\rho\equiv\gamma^{ab}{K_{ab}}^{\rho}$. Below, we provide explicit examples of charged dilatonic branes that satisfy these requirements.

\section{Measurement of Piezoelectric Moduli}
In order to measure the piezoelectric moduli from a gravitational solution we consider asymptotically flat charged dilatonic black branes in EMD theory with action
\begin{align} \label{eqn:action}
\!\!\!\!S=\frac{1}{16\pi G}\int d^{D}x \sqrt{-g} \left[ R - 2(\partial\phi)^2 - \frac{e^{-2a\phi}}{4} F^2 \right]~.
\end{align}
Ref. \cite{Armas:2011uf} measured the elastic moduli of uncharged black branes by extracting the coefficients $d^{ab\rho}$ from the stress-energy tensor measured far away from the brane horizon where the weak field approximation is valid and using Eq.~\eqref{dipole_stress}. Similarly, we extract the coefficients $p^{a\rho}$ by determining how the gauge field $A_{\mu}$ falls off at infinity and obtain $\tilde\kappa^{abc}$ using Eq.~\eqref{dipole_current}. In general, the current $\hat{J}^{\mu}$ and the gauge field are related through the equation of motion
\beq
\nabla_{\nu} \left( e^{-2a\phi} F^{\mu\nu} \right) = 16\pi G \hat{J}^{\mu} \,,
\eeq
which in Lorenz gauge, $\nabla^{\mu} A_{\mu} = 0$, and for asymptotically flat space-times gives rise to the linearized equation
\begin{equation} \label{eqn:eom}
	e^{-2a\phi_0} 
	\partial^{\nu}\partial_{\nu} A^{\mu} = - 16\pi G \hat{J}^{\mu} \,,
\end{equation}
where $\phi_0$ is the value of the dilaton far away from the brane horizon. The dipole term $p^{a\rho}$ can then be obtained by introducing the expansion \eqref{current} into Eq.~\eqref{eqn:eom}.

%%%%%%%%%%%%%%%%%%%%%%%%%%%%%%%%%%%%%%%%%%%%%%%%

\subsection*{Charged dilatonic black strings}
We now focus on a large class of asymptotically flat charged dilatonic black strings carrying 0-brane charge which can be obtained by performing an uplift of the neutral black string solution in $D=n+4$ dimensions, followed by a boost with rapidity $\alpha$ and a Kaluza-Klein (KK) reduction along a Killing direction. The resulting metric is given by \cite{Caldarelli:2010xz}
\begin{equation} \label{dsstring}
	ds^2 = -\frac{f}{h^A} dt^2 + h^B \left( \frac{dr^2}{f} + r^2d\Omega^2_{(n+1)} + dz^2 \right) \,,
\end{equation}
with $f(r) = 1 - \frac{r_0^n}{r^n}$ and $h(r) = 1 + \frac{r_0^n}{r^n} \sinh^2\alpha$, and the gauge and dilaton fields read 
\begin{eqnarray} 
	A_{t}(r) &=& -\sqrt{N} \frac{r_0^n}{r^n h(r)} \sinh\alpha  \cosh\alpha~, \\
	\phi(r) &=& -\frac{1}{4}N a \log h(r)~ . \label{ap}
\end{eqnarray}
This solution generating technique leaves the horizon regular and yields a specific value for the dilaton coupling $a^2 = \frac{2(n+3)}{n+2}$ as well as parameters  $A = \frac{n+1}{n+2},~ B = \frac{1}{n+2}$ and $N=A+B=1$.
After KK reduction, the rapidity $\alpha$ gains the interpretation of a charge parameter. The bent version of \eqref{dsstring}-\eqref{ap} can be obtained in a similar fashion using its neutral counterpart \cite{Emparan:2007wm} as a seed solution. Since we only need to know how the dipole corrections to the fields $g_{\mu\nu},A_{\mu},~\phi$ behave at infinity we focus on the large $r$-asymptotics. For use below we recall the decomposition of the metric components of the neutral bent black string given in \cite{Armas:2011uf}, 
\beq \label{ds_decomp}
g_{\mu\nu} = \eta_{\mu\nu} + h^{(M)}_{\mu\nu} + h^{(D)}_{\mu\nu} + \mathcal{O}\left( r^{-n-2} \right) \, ,
\eeq
where $h^{(M)}_{\mu\nu}$ and $h^{(D)}_{\mu\nu}$ denote the monopole and dipole contributions respectively.  The dipole contribution is parametrized as $h^{(D)}_{\mu\nu} = f^{(D)}_{\mu\nu} \varepsilon \thinspace r_0^{n+1} \left( \frac{\cos\theta}{r^{n+1}} \right)$, with coefficients $f^{(D)}_{\mu\nu}$ obtained from \cite{Emparan:2007wm},
\begin{eqnarray} \label{fff}
	f^{(D)}_{tt} &=& -(n+1) \tilde k - n(n+2)\xi(n) \,, \\
	f^{(D)}_{tz} &=& -\sqrt{n+1} \tilde k ~,\quad f^{(D)}_{zz} = -\tilde k + 2 n \xi(n) \,, \\
	f^{(D)}_{rr} &=& f^{(D)}_{\Omega\Omega} = - \tilde{k} - (n+4)\xi(n) \, ~.
\end{eqnarray}
where $\tilde k$ is the residual gauge freedom associated with the 'extra symmetry 2' and
\beq
\xi(n)=-\frac{2^{-\frac{n+4}{n}}}{n+1}\frac{\Gamma(\frac{2n+1}{n})\Gamma(-\frac{n+2}{n})}{\Gamma(-\frac{n+1}{n})\Gamma(\frac{n+2}{2n})}~.
\eeq
Turning to the charged case, we adopt a similar decomposition for the gauge field 
\beq
A_{\mu} = A^{(M)}_{\mu} + A^{(D)}_{\mu} + \mathcal{O}\left( r^{-n-2} \right) \, .
\eeq
Defining the asymptotic coefficients $a_{\mu}^{(D)}$ of the gauge field by $A^{(D)}_{\mu} = a^{(D)}_{\mu} \varepsilon \thinspace r_0^{n+1} \left( \frac{\cos\theta}{r^{n+1}} \right)$, one then finds after KK reduction,
\beq
\begin{split}
a^{(D)}_{t} &= \sinh\alpha_k \cosh\alpha_k \thinspace f^{(D)}_{tt} \\
a^{(D)}_{z} &= \sinh \alpha_k \thinspace f^{(D)}_{tz} \, ,
\end{split}
\eeq
where $n\sinh^2\alpha=(n+1)\sinh^2\alpha_k$. Using this on the left hand side of Eq.~\eqref{eqn:eom} together with the expansion \eqref{current} and the fact that $\phi_0=0$ yields
\begin{equation} %\partial^{\mu}\partial_{\mu}
	\nabla^2_{\perp} A^{(D)}_{\nu} = 16\pi G {p_{\nu}}^{r_{\perp}}\partial_{r_{\perp}} \delta^{(n+2)}(r) \,,
\end{equation}
where the Laplacian operator is taken along transverse directions to the worldvolume and $r_{\perp} = r\cos\theta$. For the given configuration at hand one finds the electric dipole moment
\begin{eqnarray} \label{eqn:dipoleCurrent}
	{p_a}^{r_{\perp}} &=& \frac{\Omega_{(n+1)} r_0^{n+2}}{16 \pi G R} a^{(D)}_a \,, \quad a=t,z \,.
\end{eqnarray}
The dipole moment of stress-energy tensor $d^{ab\rho}$ can also be obtained using the expansion \eqref{stress_tensor} and the linearized equation
\begin{equation} %\partial^{\mu}\partial_{\mu}
	\nabla^2_{\perp} \bar{h}^{(D)}_{\mu\nu} = 16\pi G  d_{\mu\nu}^{\phantom{\mu\nu}r_{\perp}} \partial_{r_{\perp}} \delta^{(n+2)}(r) \,,
\end{equation}
where the dipole perturbation $\bar{h}_{\mu\nu}^{(D)}$ of the bent charged black string is defined in analogy with \eqref{ds_decomp}. This leads to the components
\begin{eqnarray} \label{eqn:dipoleElastic}
	d_{tt}^{\phantom{tt}r_{\perp}} &=& \frac{\Omega_{(n+1)} r_0^{n+2}}{16 \pi G R}  \left(\cosh^2\alpha_k f^{(D)}_{tt} + f^{(D)}_{\Omega\Omega} \right) \,, \\
	d_{tz}^{\phantom{tz}r_{\perp}} &=& \frac{\Omega_{(n+1)} r_0^{n+2}}{16 \pi G R} \left( \cosh\alpha_k f^{(D)}_{tz} \right) \,, \\
	d_{zz}^{\phantom{zz}r_{\perp}} &=& \frac{\Omega_{(n+1)} r_0^{n+2}}{16 \pi G R} \left( f^{(D)}_{zz} - f^{(D)}_{\Omega\Omega} \right)\,,
\end{eqnarray}
expressed in terms of the asymptotic coefficients \eqref{fff} of the neutral bent black string solution.

\subsection*{Response coefficients and corrections to black rings}
The leading order effective worldvolume stress-energy tensor is of the perfect fluid form \cite{Emparan:2011hg, Caldarelli:2010xz},
\begin{equation}
	T^{ab}_{(0)} = \frac{\Omega_{(n+1)}}{16 \pi G} r_0^n \left( n(1 + N\sinh^2\alpha)u^au^b - \eta^{ab} \right) \,,
\end{equation}
where $u^{a}$ is the local boost on the string. Using this in the second equation of \eqref{eom_blackstring} one finds the leading order critical boost $u_a = [\cosh\beta_{(0)}, \sinh\beta_{(0)}]$ with $\sinh^2\beta_{(0)} = (n\cosh^2\alpha)^{-1}$. To lowest order in $\varepsilon$ the electric current takes the form $J^{a}_{(0)} = \mathcal{Q} u^{a}$ with $\mathcal{Q} = \frac{\Omega_{(n+1)}}{16 \pi G} n\sqrt{N} r_0^n \sinh\alpha\cosh\alpha$ being the charge density to the same order.  

The piezoelectric moduli are then obtained from Eqs.~\eqref{dipole_current} and \eqref{eqn:dipoleCurrent} as
\begin{eqnarray} \label{PZ}
	\tilde \kappa^{tzz} &=&  \left( \tilde k + \frac{n(n+2)}{n+1} \xi(n) \right) r_0^{2} J^{t}_{(0)} \,, \\
	\tilde \kappa^{zzz} &=& \tilde k r_0^{2} J^{z}_{(0)} \,, \label{PZ1}
\end{eqnarray}
expressed in terms of the critical current. Similarly, the elastic moduli can be obtained using Eqs.~\eqref{dipole_stress} and \eqref{eqn:dipoleElastic}
\begin{eqnarray} \label{YM}
	&&\tilde Y^{ttzz} = \tilde k r_0^{2} T^{tt}_{(0)} + \frac{\Omega_{(n+1)} r_0^{n+2}}{16 \pi G} \left(\frac{n^2(n+2)}{n+1} \sinh^2\alpha+ n^2+3n+4\right) \thinspace \xi(n) \,, \\
	&&\tilde Y^{tzzz} = \tilde k r_0^{2} T^{tz}_{(0)} \,, \\
	&&\tilde Y^{zzzz} = - \frac{\Omega_{(n+1)} r_0^{n+2}}{16 \pi G} \left( 3n + 4 \right) \xi(n) \,, \label{YM1}
\end{eqnarray}
expressed in terms of the critical stress-energy tensor. The neutral case obtained in \cite{Armas:2011uf} is reproduced when $\alpha$ is taken to zero. Using Eqs.~\eqref{eom_blackstring} we can compute the correction to the critical boost of a thin charged dilatonic black ring by bending the charged dilatonic black string described by the response coefficients \eqref{PZ}- \eqref{PZ1} and \eqref{YM}-\eqref{YM1}. This yields:
\beq
\text{sinh}^{2}\beta=(n\cosh^2\alpha)^{-1}\left(1+\varepsilon^2(3n+4)\xi(n)\right)~,
\eeq
and constitute a prediction for $n>2$ where the probe approximation is valid.

%%%%%%%%%%%%%%%%%%%%%%%%%%%%%%%%

\section{Discussion}
We have shown the existence of a new response coefficient of black branes making a solid connection between the physics of piezoelectrics and gravity. This new effect can be intuitively understood if one imagines slightly curving a charged black string of finite thickness inducing a higher concentration of charged black material in the inner surface and a depletion in the outer surface. A varying concentration of matter due to the compression and stretching of the material on opposite sides induces a bending moment of dipolar character as in classical Hookean elasticity theory and, since the matter is charged, it also induces an electric dipole moment that describes the response of the charged string to the mechanical stress. Electric fields induced by mechanical stresses are the basic feature of piezoelectrics and their behavior is governed by the physics of electroelastic materials.
As an explicit example,  we measured the elastic and piezoelectric moduli for charged dilatonic black strings. The same procedure can be applied to the $p$-branes of \cite{Camps:2012hw} charged under higher form fields and will be presented elsewhere \cite{Armas:2012}. It would be interesting to explore the physical interpretation of these response coefficients in the context of AdS/CFT, which would require obtaining the bent metric of a $D3$-brane. We expect further finite thickness effects due to the coupling to the 5-form flux, namely, an extra contribution to the dipole electric (magnetic) moment would appear which would allow us to measure electric (magnetic) susceptibilities. Examining the effect of Chern-Simons terms on the response coefficients computed here would also be interesting, in part due to the relation of these terms to the anomaly \cite{Erdmenger:2008rm, Banerjee:2008th, Son:2009tf} via the gauge-gravity correspondence.  A more formal development of the electroelasticity of black branes could be achieved by obtaining the metric of a bent black string to second order in $\varepsilon$ and we will leave that for future work. In that context, it will be interesting to investigate whether relations of the type in Eqs. \eqref{dipole_stress} and \eqref{dipole_current}, and generalizations thereof, can be proven using general covariance and the laws of thermodynamics.

\section*{Acknowledgements}
JA and NO wish to thank the KITP for hospitality during the program ``Bits, branes and black holes'', and acknowledge that this research was supported in part by the National Science Foundation under Grant No. NSF PHY11-25915. The work of NO is supported in part by the Danish National Research Foundation project ``Black holes and their role in quantum gravity''. The work of JA was funded by FCT Portugal grant SFRH/BD/45893/2008.

\addcontentsline{toc}{section}{References}
%\small
\providecommand{\href}[2]{#2}\begingroup\raggedright\endgroup

\end{document}